\documentclass[referee]{aa}
\usepackage{graphicx}
\usepackage{psfig}
\usepackage{amsfonts}

\begin{document}

\title{ Globular Clusters and the Horizontal Branch }
\author{Z.\,Kadla\inst{1} }

\offprints{Z.\,Kadla}

 \institute{Main Astronomical Observatory of
the Russian Academy of Sciences at Pulkovo, 196140
Saint-Petersburg, Russia (kadla@gao.spb.ru)}

\date{Received ; accepted }

\abstract{An analysis of available Galactic globular cluster data
on metallicity, ages, the distribution of stars on the horizontal
branch and the distribution of the periods of RR0 Lyr variables
showed that:

1. The latter two characteristics can be used for determining the
Oosterhoff type.

2. The metallicities of the observed OoI and OoII clusters overlap
in a narrow interval -1.65$<$[Fe/H]$<$1.52.

3. The OoII clusters are characterized by R-B$<$0, a statistically
significant relation (r=0.97) between the relative number of B(BS) and
V(VS) stars on the horizontal branch.

4. The OoI clusters divide into two groups: R-B$<$0 and R-B$>$0
with a relation between B and R (r=0.90). The former clusters do
not show a relation between VS and RS stars and the latter a
relation between VS and BS stars on the horizontal branch.

5. The oldest clusters are characterized by  BS close to 1 and the
youngest by R close to 1.

 }

\maketitle

\section{Introduction}

       The existence of two types of globular clusters (GC) OoI and OoII
   was first noted by Oosterhoff (1941), the GC types being represented by different
period-amplitude relations. However later Belserene (1954) showed
that the RR Lyrae variables in $\omega$\,Cen, M\,5 and M\,3 are
divided into two sequences and the clusters differ from each other
in the relative population of the sequences. From an analysis of
the OoII cluster NGC\,5139 she found six variables belonging to
OoI type clusters. In a detailed
   investigation of the Oo1 type cluster NGC\,5272 Szeidl (1965) discovered
   six RRab stars belonging to OoII type clusters. Further studies revealed
   a considerable dispersion of points in the RRab period-amplitude diagrams
   of globular clusters. The mean period of RR0 (previous designation RRab,
   Clement et al., 2001) variables belonging to OoI GCs are characterized by
   0.5$<$mP0$<$0.6 and the OoII by mP0$>$0.6 (Clement et al., 2001). The clusters
   differ in horizontal branch morphology. Clusters with R-B$>$0 belong to the
   OoI type GC (Kadla \& Geraschenko, 1984). Using available observational data
   the mP0 classification is applied to clusters with the number of RR0
   variables $>$2 and determined distribution of stars (B, V, R) on the
   horizontal branch.

\section{Observational Data}

       In the present study we used data from the following two catalogues.
   The "Catalog of Parameters for Milky Way Globular Clusters" compiled by
   Harris (2003) which contains data for 150 GC, 148 of which have determined
   metallicities (exceptions BH 176 and 2MS-GC02). The data in the catalogue
   "Variable Stars in Globular Clusters" compiled by Clement et al.
   (2001) were summarized by the authors in tables 1 and 2.

      The following characteristics of the HB were used for analysis:

   mP0 - the mean period of RR0 variables;

   nRR0 - the number of RR0 variables used for determining mP0;

   B - the number of stars on the blue HB (BHB);

   V - the number of RR Lyr variables;

   R - the number of stars on the red HB (RHB);

   HBR - the horizontal-branch ratio (B-R)/(B+V+R);

   [Fe/H] - metallicity

       The following were computed for individual clusters: (B+V+R) = S,
   B/S = BS, V/S = VS, R/S = RS (the relative number of stars on the
   BHB, VHB and RHB).

    Values of B, V and R with references to the investigators of the
   individual clusters are given in the papers by Lee et al. (LDZ,
   1994, 83 GC, table 1), Preston et al. (PSB, 1991, 44 GC,
   table 4) and Brocato et al. (BBMP, 1996, 9 GC, table 2). The LDZ list
   includes 42 of the clusters in PSB and 8 in BBMP.
   The relative
   number of stars on the HB computed from the LDZ and PSB lists are in
   excellent agreement (fig.1). In the following L denotes LDZ, P - PSB and
   BR - BBMP.

          BSP=-0.023+1.013BSL; n=42; r=0.99

          VSP=-0.002+1.048VSL; n=42; r=0.96

          RSP= 0.010+0.999RSL; n=42; r=0.99

\begin{figure} [h]
\centering{
\vbox{\psfig{figure=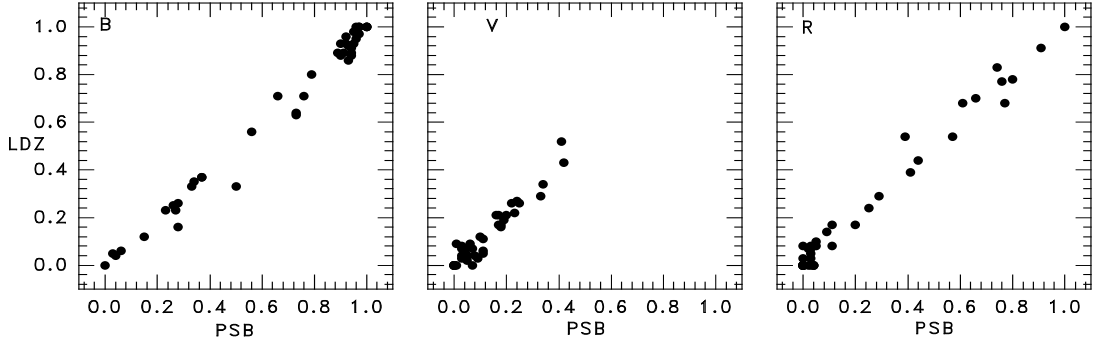,width=14.5cm,height=4.5cm}}\par}
\caption []{ A comparison of the relative number of stars (BS, VS,
RS) on the HB in the LDZ and PSB lists.}
\end{figure}

   The relations determined from a similar comparison of the BBMP list
   have a lower correlation coefficient (r).

         BSBR=-0.159+1.144BSL; n=9; r=0.94

         VSBR= 0.043+0.926VSL; n=9; r=0.85

         RSBR= 0.025+0.933RSL; n=9; r=0.82

       There are altogether 72 clusters with determined B, V and R within the
   metallicity interval -2.29$<$[Fe/H]$<$-0.90. The mP0 classification applied
   to GCs with nRR0$>$2 and [Fe/H]$<$-0.90 indicates that the OoII GCs lie within
   the metallicity interval -2.29$<$[Fe/H]$<$-1.58 and the OoI -1.83$<$[Fe/H]$<$-0.95
   (fig. 2). With the exception of NGC\,4147 (-1.83) the two sequences overlap
   in the interval -1.65$<$[Fe/H]$<$-1.57 (fig.2). The maximum values for OoII
   clusters are -1.64, -1.62, -1.58 (mean -1.61). With the exception of
   NGC\,4147 (-1.83) the minimum values for OoI GCs are -1.63, -1.59,-1.58
   (mean -1.60). The mean period of RR0 variables and metallicity are plotted
   in fig. 2.

\begin{figure}[h]
\centering{
\vbox{\psfig{figure=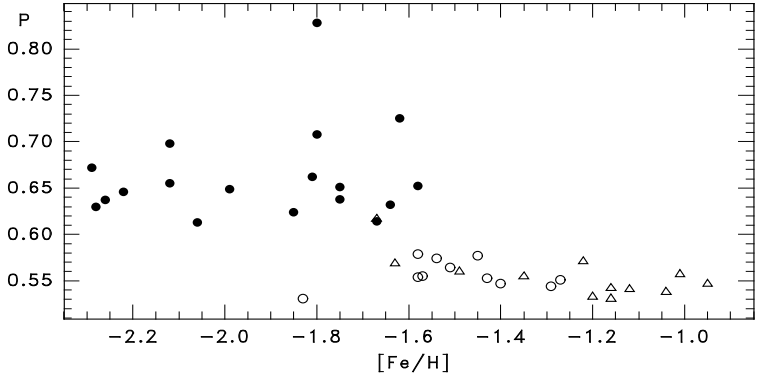,width=10cm,height=5cm}}\par}
\caption[]{ Metallicity and the mean period of RR0 variables.
$\bullet$ - OoII type cluster. $\circ$ - OoI with $R-B<$0,
$\triangle$ - OoI with $R-B>$0.}
\end{figure}

       According to the catalogue Clement et al. (2001) RR0 variables have
   been discovered in 73 GCs (about one-half of the known GC) and for
   54 (4 with [Fe/H]$>$-0.90) of these  have determined B and R .

\section{OoII globular clusters}

       The adopted mP0 classification indicates that there are 19 OoII GC
   with nRR0$>$2 in the LDZ list,.15 in PSB and 2 in BBMP. R-B$<$0 for all the OoII
   type clusters. The cluster Rup 106 with mP0 = 0.617 is an exception. As
   noted by Clement et al. (2001) the P-A relation for the GC is similar to that of
   M3 (Kaluzny et al., 1995) and should be classified as OoI. It is also an exception
   as according to R-B$>$0 it belongs to OoI. The P-A relations for Rup 106,
   NGC\,7089 (OoII) and NGC\,3201 (OoI) with metallicities -1.67. -1.62 and
   -1.58 respectively are plotted in fig 3. Evidently the upper value of mP0
   for OoI clusters should be increased to 0.613.

\begin{figure}[h]
\centering{
\vbox{\psfig{figure=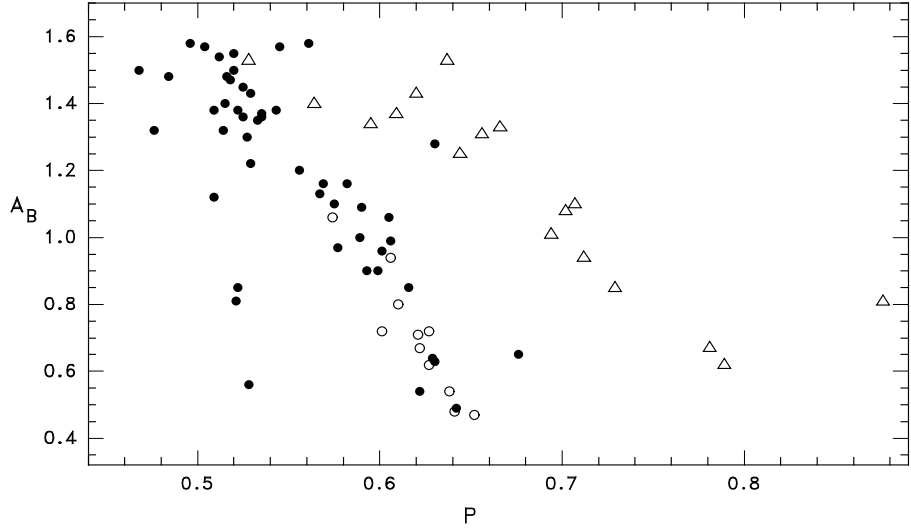,width=12.0cm,height=7cm}}\par}
\caption[]{The P-A relation for Rup 106 ($\circ$), NGC\,7089
($\triangle$) and NGC\,3201 ($\bullet$) }
\end{figure}

       Excluding Rup 106 the following relations derived for the OoII GC
 are shown in fig.4.

          VS=0.752-0.756BS=0.754(1-BS); n=35; r=-0.97

          RS=0.245-0.240BS=0.242(1-BS); n=35; r=-0.79

\begin{figure}[h]
\centering{
\vbox{\psfig{figure=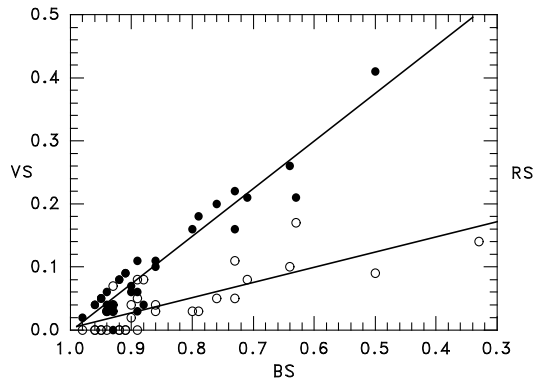,width=7cm,height=5cm}}\par}
\caption[]{The relation between RS and BS for the OoII clusters,
$\bullet$ - VS, $\circ$ -RS.}
\end{figure}

   Among the observed GCs the young cluster NGC\,4590 has the largest
   value of VS = 0.41 (RS = 0.09). Note that the observed relative number of
   R stars in the OoII clusters does not exceed 0.17 and R-B$<$0 for all the
   OoII type clusters.

       In the PSB list there are 43 GC with [Fe/H]$<$-0.90, and one (NGC\,104)
   with [Fe/H]=-0.76. According to the adopted mP0 classification 23 belong
   to OoII and 17 to OoI. The 15 OoII GC with nRR0$>$2 also show a strong
   correlation between VS and BS .

         VS=0.729-0.734BS=0.731(1-BS); n=15; r=-0.98

         RS=0.264-0.259BS=0.261(1-BS); n=15; r=-0.82

      In the metallicity interval -1.57$<$[Fe/H]$<$-1.51 there are 3 OoII clusters
 with nRR0$<$3 and two with a HB distribution which indicates that they
    belong to this group. If these clusters are taken into account the
    maximum metallicity average for the OoII GCs equals -1.52.

     In the list compiled by (BBMP) there are 9 GC,
    4 of which belong to OoII and 5 to OoI. However only two conform to
    the adopted at present criteria for OoII clusters. Values of mP0
    for NGC\,4372 and NGC\,6218 have not been determined.

\section{OoI globular clusters}

       The OoII GCs are characterized by $R-B<$0, while the OoI GCs include
    clusters with $R-B<$0 and $R-B>$0. There are 21 OoI GCs with nRRO$>$2 and
    [Fe/H]$<$-0.95 in the LDZ list, 15 in PSB and 3 in BBMP (table 1).

\begin{table}[h]
\caption[ ]{}
\begin{tabular}{cccc}
\hline
\noalign{\smallskip}
 $R-B$ & LDZ & PSB & BBMP \\
  \hline
  \noalign{\smallskip}
 $R-B<0$  & 10 & 6 & 3 \\
 $R-B>0$  & 11 & 9 & 0 \\
  \hline
 \noalign{\smallskip}
\end{tabular}
\end{table}

 Only BS and RS of the considered clusters are correlated (fig 5).

         BSL=-0.676-0.745RSL; n=21; r=-0.88

         BSP=0.693-0.772RSP; n=15; r=-0.91

       In the LDZ list there are also 14 GC with [Fe/H]$>$-0.90, which
       according to their B and R values (BS=0, RS=1) belong to OoI.

\begin{figure}[h]
\centering{
\vbox{\psfig{figure=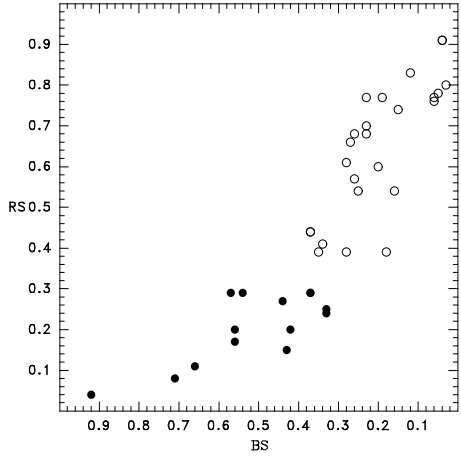,width=6cm,height=6cm}}\par}
\caption[]{The relation between RS and BS for the OoI GC.
$\bullet$ - clusters with $R-B<$0, $\circ$ - clusters with
R-B$>$0.}
\end{figure}

If the OoI GCs with $R-B<$0 and $R-B>$0 are considered separately
a relationship between VS and RS is absent for the former and VS
and BS for the latter.

Here we note that the data for NGC\,6715 in the PSB list
(BS-RS=0.21) differs from that given by Harris (HBR=0.83). If the
latter is correct BS=0.90, $VS$=0.09, RS=0.01. (PSB data
respectively 0.42. 0.37, 0.21). NGC\,6715 is not in the LDZ list.
There are several GC which fall out of the general scheme. Before
reaching a final conclusion the data for such clusters should be
checked.

\section{Cluster Ages}

The ages (BMA) of 65 GCs based on the relative ages determined by
Buonanno et al. (1998) are given in the paper by Borkova \&
Marsakov (2000, BMA) who analysed the age lists published before
1999. Details on the method and lists used are given in their
paper.

The latest homogeneous ages (SWA) for 55 GC were determined by
Salaris \& Weiss (2002, SWA). There are altogether 48 GC with
determined ages both in SWA (column 4 table 4) and BMA.The
distribution of stars on the HB  has not been determinated for
four of these clusters (NGC\,6652, IC\,4499, Arp\,2 and Terz\,7).
There is poor agreement between the two sets of age
determinations:

          SWA = 1.024 + 0.714BMA;  r = 0,69; n = 48

The differences between the two sets of age estimates E(Gyr) = BMA-SWA
are shown in fig. 6

\begin{figure}[h]
\centering{
\vbox{\psfig{figure=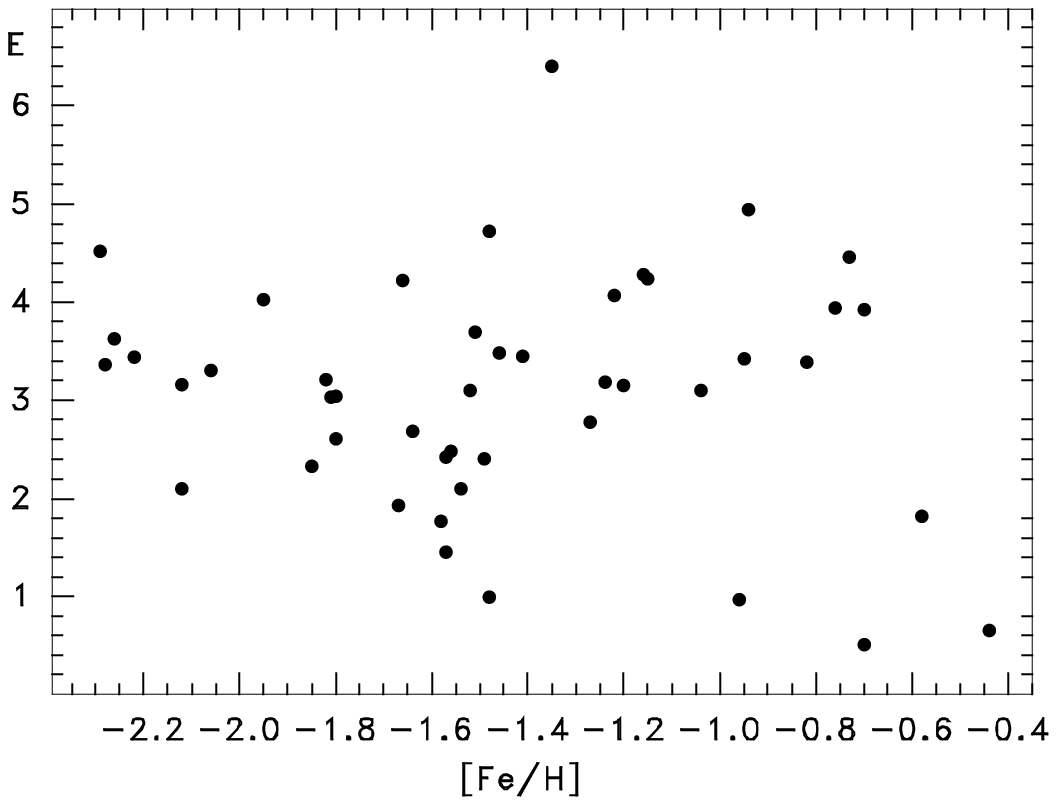,width=10cm,height=7cm}}\par}
\caption[]{E(Gyr) = BMA-SWA}
\end{figure}

The metallicity interval -1.85 to -1.40 stands out in that there
are 20 clusters with E$<$3. There is a much closer agreement if
clusters with 3$<$E$<$4.95 and E$>$3 are considered separately

        SWA=-6.253+1.176BMA;  r=0.94;  28; E$>$3.

        SWA= 1.949+0.720BMA;   r=0.93; 20;  E$<$3.

 The number of clusters in each SWA age group are given in table 2 and fig. 7.

\begin{table}[h]
\caption[ ]{}
\begin{tabular}{cccccccc}
\hline \noalign{\smallskip}
 SWA & OoI &  OoI &  OoII &   Sum & OoI & OoI &  OoII \\
   &$R-B>$0&$R-B<$0&  &  &$R-B>$0&$R-B<$0\\
  \hline
  \noalign{\smallskip}
6.4-10.6 &   12 +(6)& 0 &  0 &  18 & 0.33 &   0.00 &   0.00 \\
10.7-11.9 & 5 & 3 &  7 &  15 & 0.09 &   0.05 &   0.13 \\
11.9-12.9& 1 & 2 & 19 & 22 & 0.02 & 0.04 & 0.35  \\
 \hline
 \noalign{\smallskip}
 & 18 +(6) & 5 &  26 & 55 & 0.44 & 0.09 & 0.48\\
 \hline
 \noalign{\smallskip}
\end{tabular}
\end{table}

\begin{figure}
\centering{
\vbox{\psfig{figure=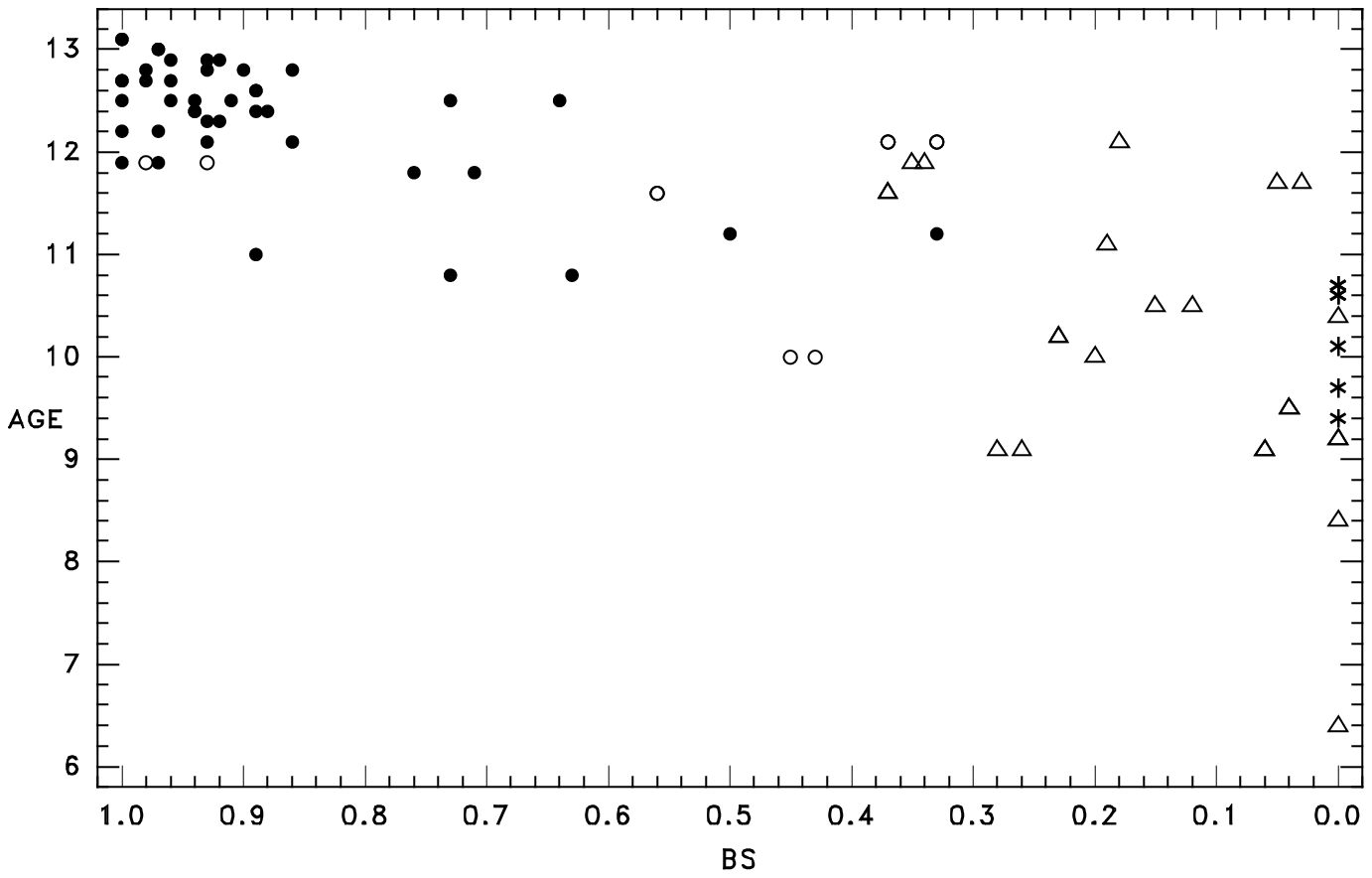,width=15cm,height=8cm}}\par}
\caption[]{ The relation between $BS$ and SWA.  $\bullet$ - OoII
type cluster. $\circ$ - OoI with $R-B<$0, $\triangle$ - OoI with
$R-B>$0, $*$ - clusters with [Fe/H]$>$-0.9.}
\end{figure}

The numbers in brackets denote the number of clusters with
[Fe/H]$>$-0.90. It follows that the youngest clusters belong to
OoI with $R-B>$0 and the oldest to OoII. The young OoII clusters
are the same age as the old OoI GC. The relation between BS and
SWA for OoII clusters shows that BS increases with age.

     SWA =10.556 + 2.177BS;  n=19; r=0.89

The data for BMA are given in table 3 and fig.8.

\begin{table}[h]
\caption[ ]{}
\begin{tabular}{cccccccc}
\hline \noalign{\smallskip}
 BMA & OoI & OoI & OoII & Sum& OoI & OoI & OoII \\
      &$R-B>$0&$R-B<$0&  &  &$R-B>$0&$R-B<$0\\
  \hline
  \noalign{\smallskip}
9.0-14.0& 12+(6)& 0 & 0 & 18 & 0.33 & 0.00 & 0.00   \\ 14.0-15.0 &
5 & 3 & 7 & 15 & 0.09 & 0.05 & 0.13   \\ 15.0-16.6 & 1 & 2 & 19&
22 & 0.02 & 0.04 & 0.35   \\
 \hline
 \noalign{\smallskip}
 & 19+(11) & 9 &  25 & 64 & 0.47& 0.14& 0.39\\
 \hline
 \noalign{\smallskip}
\end{tabular}
\end{table}

\begin{figure}[h]
\centering{
 \vbox{\psfig{figure=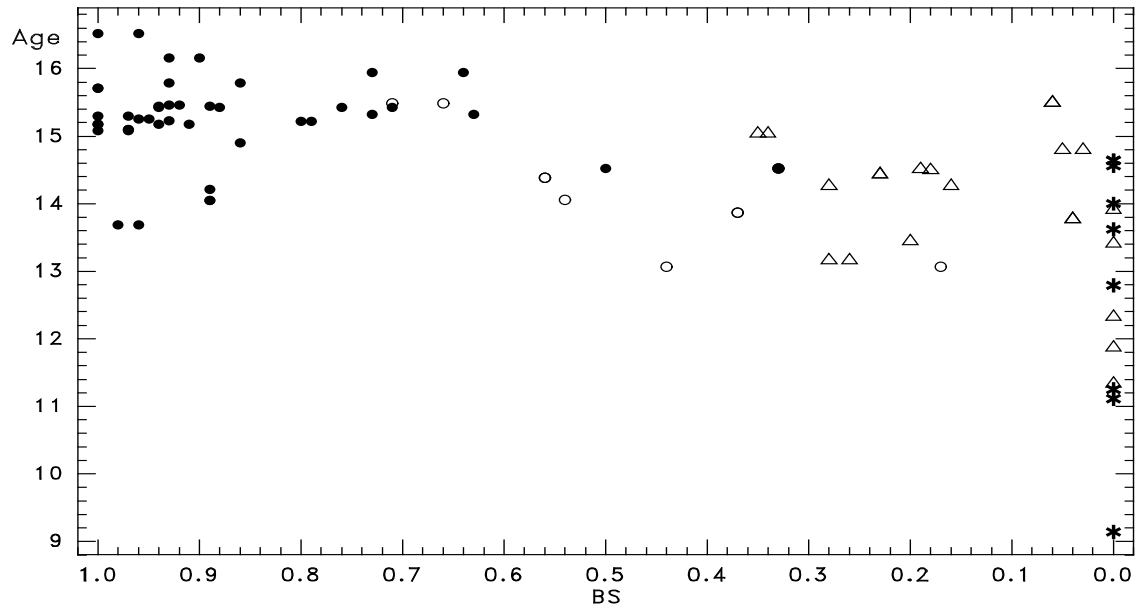,width=15cm,height=8cm}}\par}
\caption[]{ The relation between BS and BMA. $\bullet$ - OoII type
cluster. $\circ$ - OoI with $R-B<$0, $\triangle$ - OoI with
$R-B>$0, $*$ - clusters with [Fe/H]$>$-0.9. }
\end{figure}

The cluster NGC\,6626, age 19.1 (BMA), has not been included.
According to mP0 = 0.577, nRR0 = 8, it belongs to OoI. However
with BS = 0.92 and VS = 0.04 it can be classified as OoII.
 The data used for analysis are given in table 4.

\section{Conclusions}

The mP0 classification and the distribution of stars on the HB are
indications of the Oo type. The Oosterhoff type clusters OoI are
characterized by the mean period of RR0 variables
0.5$<$mP0$<$0.613 and the type OoII by mP0$>$0.6. The mP0
classification indicates that the OoII clusters lie within the
metallicity interval -2.29$<[Fe/H]<$-1.58 and the OoI
-1.83$<[Fe/H]<$-0.95. With the exception of NGC\,4147 (-1.83) the
two sequences overlap in the interval -1.65$<[Fe/H]<$-1.57 (table
1 and fig.2). The OoII GCs are characterized by $R-B<$0, while the
OoI GCs include clusters with $R-B<$0 and $R-B>$0.

The BS and VS relation for the OoII clusters reveals a strong
correlation between the two characteristics.

        The oldest clusters belong to OoII with BS close to 1 and the youngest
    to OoI $R-B>0$ with RS=1. The characteristic R/B decreases with age.

        The 14 clusters with [Fe/H]$>$-0.90 and determined distribution of
    stars on the HB belong to OoI.

                                         `

\begin{table}[p]
\caption[ ]{}
\begin{tabular}{lccccccccccccccc}
\hline \noalign{\smallskip}
 Cl& FeH & SWAA &BMA & $BS$L& $VS$L& $RS$L& SumL & $BS$P & $VS$P& $RS$P& SumP& mP0& RR0\\
  \hline
  \noalign{\smallskip}
104 &    -0.76&   10.7& 14.64& 0.00& 0.00& 1.00&  77& 0.00& 0.00&
1.00& 336& 0.737&  1\\ 288 &    -1.24&   11.9& 15.08& 0.97& 0.01&
0.02& 102& 1.00& 0.00& 0.00&  95& 0.678&  1\\ 362 &    -1.16&
9.5& 13.78& 0.04& 0.05& 0.91&  78& 0.04& 0.04& 0.91&  92& 0.542&
7\\ Erid &   -1.46&    8.4& 11.88& 0.00& 0.00& 1.00&  12  \\ 1851
&   -1.22&    9.1& 13.17& 0.28& 0.11& 0.61& 103& 0.26& 0.06& 0.68&
96& 0.571& 21\\ 1904 &   -1.57&   12.6& 14.05& 0.89& 0.11& 0.00&
35& 0.89& 0.11& 0.00&  35& 0.685&  2\\ 2298 &   -1.85&   12.9&
15.23& 0.93& 0.07& 0.00&  15&     &     &     &    & 0.649&  1\\
2419 &   -2.12&   12.8& 14.90& 0.86& 0.11& 0.03& 107&     &     &
&    & 0.655& 24\\ 2808 &   -1.15&   10.2& 14.44& 0.23& 0.01&
0.77& 265& 0.23& 0.09& 0.68&  56& 0.539&  1\\ Pal 3 &  -1.66&
9.2& 13.42& 0.00& 0.18& 0.82&  17  \\ 3201  &  -1.58&   12.1&
13.87& 0.37& 0.34& 0.29& 176& 0.37& 0.34& 0.29& 176& 0.554& 72\\
Pal 4 &  -1.48&    9.2& 13.92& 0.00& 0.00& 1.00&  20  \\ 4147  &
-1.83&       & 15.49& 0.66& 0.23& 0.11&  62& 0.71& 0.22& 0.08&
51& 0.531&  4\\ 4372  &  -2.09&       &      & 1.00& 0.00& 0.00&
76  \\ Rup 106& -1.67&   10.4& 12.33& 0.00& 0.18& 0.82&  50&     &
&     &    & 0.617& 13\\ 4590 &   -2.06&   11.2& 14.52& 0.50&
0.41& 0.09&  44& 0.33& 0.52& 0.14&  21& 0.613& 13\\ 4833 &
-1.80&       &      & 0.95& 0.05& 0.00& 175& 0.93& 0.04& 0.03&
180& 0.708&  7\\ 5024 &   -1.99&       & 15.22& 0.79& 0.18& 0.03&
176& 0.80& 0.16& 0.03& 188& 0.649& 29\\ 5053 &   -2.29&   10.8&
15.32& 0.73& 0.16& 0.11&  44& 0.63& 0.21& 0.17&  48& 0.672&  5\\
5272 &   -1.57&   12.1& 14.52& 0.33& 0.42& 0.25& 226& 0.33& 0.43&
0.24& 240& 0.555&145\\ 5286 &   -1.67&       &      & 0.91& 0.09&
0.00&  35& 0.89& 0.03& 0.08&  36& 0.614&  8\\ 5466 &   -2.22&
12.5& 15.94& 0.73& 0.22& 0.05&  77& 0.64& 0.26& 0.10&  80& 0.646&
13\\ 5694 &   -1.86&       &      & 1.00& 0.00& 0.00&  30  \\ 5824
&   -1.85&       &      & 0.86& 0.10& 0.04&  50&     &     &     &
& 0.624&  7\\ Pal 5&   -1.43&   10.0& 13.45& 0.20& 0.20& 0.60&  20
\\ 5897 &   -1.80&   12.4& 15.44& 0.94& 0.04& 0.03& 108& 0.89&
0.06& 0.05&  96& 0.828&  3\\ 5904 &   -1.27&   11.6& 14.38& 0.56&
0.24& 0.20& 164& 0.56& 0.27& 0.17& 163& 0.551& 91\\ 5927 & -0.37&
&  9.14& 0.00& 0.00& 1.00&  60  \\ 5986 &   -1.58& &      & 0.95&
0.05& 0.00&  56& 0.98& 0.02& 0.00&  50& 0.652&  7\\ Pal 14& -1.52&
&      & 0.00& 0.00& 1.00&  16  \\ 6093 & -1.75& 12.9&      &
0.92& 0.08& 0.00&  37& 0.96& 0.04& 0.00& 45& 0.651& 4\\ 6101 &
-1.82&   11.0& 14.21& 0.89& 0.06& 0.05& 83  \\ 6121 & -1.20& 11.9&
15.05& 0.34& 0.25& 0.41& 153& 0.35& 0.26& 0.39& 141& 0.533& 31\\
6144 &   -1.75&       &      & 1.00& 0.00& 0.00& 22  \\ 6171 &
-1.04&   11.7& 14.80& 0.03& 0.17& 0.80&  89& 0.05& 0.17& 0.78& 99&
0.538& 15\\ 6205 &   -1.54& 13.0& 15.10& 0.97& 0.03& 0.00& 113&
0.97& 0.03& 0.00& 106& 0.750& 1\\ 6218 &   -1.48& 12.7& 13.69&
0.96& 0.00& 0.04&  73  \\ 6229 &   -1.43&       & 14.06& 0.54&
0.17& 0.29&  83&     &     &     & & 0.553& 30\\ 6235 & -1.40&
&      & 0.90& 0.10& 0.00& 30& 0.88& 0.12& 0.00& 26& 0.6  &  2\\
6254 &   -1.52&   12.2& 15.30& 0.97& 0.00& 0.03& 65& 1.00& 0.00&
0.00&  61  \\ Pal 15& -1.90&       &      & 1.00& 0.00& 0.00&  40
\\ 6266 &   -1.29& &      & 0.57& 0.14& 0.29& 110&     &     &
&    & 0.544& 62\\ 6293 &   -1.92&       & & 0.90& 0.10& 0.00& 41&
&     & &    & 0.6  &  2\\ 6316 & -0.55&       &      & 0.00&
0.00& 1.00&  60  \\ 6333 &   -1.75& & & 0.90& 0.06& 0.04& 83& & &
&    & 0.638&  8\\
\end{tabular}
\end{table}

\begin{table}
\begin{tabular}{lccccccccccccccc}
\hline \noalign{\smallskip}
 6341 &   -2.28&   12.8& 16.16&0.90& 0.07& 0.02& 134& 0.93& 0.00& 0.07& 129& 0.630& 11\\
 6342 &-0.65& &      & 0.00& 0.00& 1.00& 30  \\ 6352 & -0.70& 9.7& 13.62&
0.00& 0.00& 1.00& 30  \\ 6362 &   -0.95& 11.1& 14.52& 0.19& 0.04&
0.77& 90&     & &     &    & 0.547& 18\\ 6366 & -0.82& 9.4& 12.79&
0.00& 0.02& 0.98&  52& & &     & & 0.513& 1\\ 6397 & -1.95& 12.5&
16.52& 0.96& 0.00& 0.04&  79& 1.00& 0.00& 0.00& 139
\\ 6402 &   -1.39&       & & 0.65& 0.35& 0.00& 97&     &     &
&    & 0.564& 39\\ 6496 & -0.64& &      & 0.00& 0.00& 1.00&  35
\\ 6535 &   -1.80& 13.1& 15.71& 1.00& 0.00& 0.00&  16& 1.00& 0.00&
0.00&  15  \\ 6539 &   -0.66& &      & 0.00& 0.00& 1.00&  30
\\ 6541 & -1.83&       &      & 1.00& 0.00& 0.00&  15& 1.00& 0.00&
0.00&  75
\\ 6544 &   -1.56&       &      & 1.00& 0.00& 0.00&  38&     &
&     &    & 0.570&  1\\ 6553 &   -0.21&       & 14.00& 0.00&
0.00& 1.00&  47&     &     &     &    & 0.526&  2\\ 6584 &
-1.49&   12.1& 14.50& 0.18& 0.42& 0.39&  38&     &     &     &
& 0.560& 34\\ 6624 &   -0.44&   10.6& 11.25& 0.00& 0.00& 1.00&  22
\\ 6626 &   -1.45&       & 19.10& 0.92& 0.04& 0.04&  49& 0.93&
0.07& 0.00&  43& 0.577&  8\\ 6637 &   -0.70&   10.6& 11.11& 0.00&
0.00& 1.00&  60 \\ 6638 &   -0.99&       &      & 0.26& 0.17&
0.57&  23& 0.25& 0.21& 0.54&  24& 0.666&  1\\ 6656 &   -1.64&
12.5& 15.18& 0.94& 0.06& 0.00& 117& 0.91& 0.09& 0.00&  99& 0.632&
10\\ Pal 8&   -0.48&       &      & 0.00& 0.00& 1.00&  15 \\ 6681
&   -1.51&   11.9&      &     &     &     &    & 0.93& 0.07& 0.00&
15& 0.564&  1\\ 6712 &   -1.01&   10.5&      & 0.15& 0.11& 0.74&
94& 0.12& 0.05& 0.83& 118& 0.557&  7\\ 6715 &   -1.58&       &
&     &     &     &    & 0.42& 0.37& 0.20&  59& 0.579& 55\\ 6723 &
-1.12&   11.6&      & 0.37& 0.19& 0.44& 106& 0.37& 0.19& 0.44&
106& 0.541& 23\\ 6752 &   -1.56&   12.7& 15.18& 1.00& 0.00& 0.00&
211& 1.00& 0.00& 0.00& 196  \\ 6760 &   -0.52&       &      &
0.00& 0.00& 1.00&  30 \\ 6779 &   -1.94&   12.8&      & 0.98&
0.02& 0.00&  62&     &     &     &    & 0.906&  1\\ 6809 &
-1.81&   12.4& 15.43& 0.94& 0.03& 0.03& 102& 0.88& 0.04& 0.08&
215& 0.662&  4\\ 6838 &   -0.73&   10.1& 14.56& 0.00& 0.00& 1.00&
32 \\ 6864 &   -1.16&       &      & 0.27& 0.07& 0.66& 137& 0.23&
0.07& 0.70& 129& 0.531&  3\\ 6934 &   -1.54&   10.0&      & 0.43&
0.42& 0.15&  72&     &     &     &    & 0.574& 68\\ 6981 &
-1.40&       & 13.07& 0.44& 0.29& 0.27&  48&     &     &     &
& 0.547& 24\\ 7006 &   -1.63&       & 14.27& 0.28& 0.33& 0.39&
164& 0.16& 0.29& 0.54&  85& 0.569& 53\\ 7078 &   -2.26&   11.8&
15.43& 0.76& 0.20& 0.05& 152& 0.71& 0.21& 0.08& 188& 0.637& 39\\
7089 &   -1.62&       & 15.25& 0.96& 0.04& 0.00&  83& 0.95& 0.05&
0.00&  94& 0.725& 17\\ 7099 &   -2.12&   12.3& 15.46& 0.93& 0.03&
0.04&  67& 0.92& 0.08& 0.00&  64& 0.698&  3\\ Pal 12&  -0.94&
6.4& 11.34& 0.00& 0.00& 1.00&   7 \\ 7492 &   -1.51&   12.1&
15.79& 0.93& 0.03& 0.03&  29& 0.86& 0.07& 0.07&  28& 0.805&  1\\
 \hline
 \noalign{\smallskip}
\end{tabular}
\end{table}


\begin{thebibliography}{}

\bibitem[ ]{} {Belserene, E. P., 1954, AJ 59, 406}
\bibitem[ ]{} {Borkova T.,  Marsakov V., 2000, Astron Rep 44, 750. (BMA)}
\bibitem[ ]{} {Brocato E., Buonanno R., Malakhova Y., Piersimony A., 1996, A\&A 311,
789. (BBMP)}
\bibitem[ ]{} {Buonanno R., Corsi C.E., Pulone L., Fusi Pecci F., Bellazzini M., 1998, A\&A 333, 305}
\bibitem[ ]{} {Caretta E., Gratton R.G., Clementini G., Fusi Pecci F., 2000, ApJ 553, 215}
\bibitem[ ]{} {Clement C. M., Muzzin, A., Dufton, Q. et al. , 2001, AJ 122, 2587}
\bibitem[ ]{} {Harris W.E., 2003, AJ 112, 1487, 1996}
\bibitem[ ]{} {Kadla Z., Geraschenko A.N. , 1984, Izv. Pulkovo (Russian) 202}
\bibitem[ ]{} {Kaluzny J., Krzemieski W., Mazur B. , 1995, AJ 110, 2206}
\bibitem[ ]{} {Lee Y.-W., Demarque P., Zinn R., 1994, ApJ 243, 2481}
\bibitem[ ]{} {Oosterhoff P. Th., 1941, Leiden Ann. 17}
\bibitem[ ]{} {Preston G., Shestmanet S., Beers T., 1991, ApJ 375, 121}
\bibitem[ ]{} {Salaris M., Weiss A., 2002, A\&A 388, 492. (SWA)}


\end{thebibliography}
\end{document}